# Evaluation of the Exradin A30 Parallel Plate Ion Chamber as a Reference Dosimeter in Ultra-High Dose Rate (UHDR) Electron Beams

Mervat Alharbi[1], Kevin Liu[2], Brian Hooten[3], Shannon Holmes[3], Stefanno Alarcon-Nunez[1], Jeffrey Radtke[1], Larry DeWerd[1], Sam Beddar[2], Wesley Culberson[1], Emil Schueler[2], Ahtesham Khan[1, *]

[1]Department of Medical Physics, School of Medicine and Public Health, University of Wisconsin-Madison, Madison, WI.

[2]Department of Radiation Physics Division of Radiation Oncology, Division of Radiation Oncology, The University of Texas MD Anderson Cancer Center, Houston, Texas

[3]Standard Imaging Inc, Middleton, WI.

*Author to whom correspondence should be addressed.
Email: akhan49@wisc.edu, Work address: 1111 Highland Ave Rm B1002, Madison, WI 53705

**Running title:** Evaluation of A30 in UHDR electron beam.

**Conflict of interest**

Shannon Holmes, Larry DeWerd, and Brian Hooten are employees at Standard Imaging Inc.

**Funding statement**

This publication was supported by the National Cancer Institute of the National Institutes of Health under Topic 434−75N91022C00039 - Phase I SBIR Contract with Standard

Imaging Inc., R01CA266673, and P30 CA016672; by the University Cancer Foundation via the Institutional Research Grant program at MD Anderson Cancer Center; by the University of Texas MD Anderson Cancer Center UTHealth Graduate School of Biomedical Sciences Dr. John J. Kopchick Fellowship, Houston, TX ; by the University of Texas MD Anderson Cancer Center UTHealth Graduate School of Biomedical Sciences American Legion Auxiliary Fellowship, Houston, TX ; and by UTHealth Innovation for Cancer Prevention Research Training Program Predoctoral Fellowship (Cancer Prevention and Research Institute of Texas grant RP210042).

**Author Contribution Statement**

Mervat Alharbi contributed to the study design, performed the experiments, analyzed and interpreted the data, prepared the figures, and drafted the manuscript. Kevin Liu contributed to the study design, analyzed and interpreted the data, and drafted the manuscript. Brian Hooten contributed to preparing the device schematic and resolving the technical challenges. Stefanno Alarcon-Nunez and Jeffrey Radtke contributed to the study design and assisted with data collection. Shannon Holmes contributed to study design, research supervision, and manuscript revision. Larry DeWerd, Sam Beddar, and Emil Schuele contributed to supervision of the project. Wesley Culberson contributed to the study design, interpreted the data, and provided research supervision. Ahtesham Khan contributed to the study design, analyzed and interpreted the results, and revised the manuscript.

## Data availability statement

Research data are stored in an institutional repository and will be shared upon request to the corresponding author.

## Acknowledgments

The authors thank the students and staff of the UWMRRC for their continued support, and the UWRCL and UWADCL customers whose calibrations help support ongoing student research at the UWMRRC. Research reported in this publication was supported by the National Cancer Institute of the National Institutes of Health under Topic 434−75N91022C00039 - Phase I SBIR Contract with Standard Imaging Inc., R01CA266673, and P30 CA016672; by the University Cancer Foundation via the Institutional Research Grant program at MD Anderson Cancer Center; by the University of Texas MD Anderson Cancer Center UTHealth Graduate School of Biomedical Sciences Dr. John J. Kopchick Fellowship, Houston, TX ; by the University of Texas MD Anderson Cancer Center UTHealth Graduate School of Biomedical Sciences American Legion Auxiliary Fellowship, Houston, TX ; and by UTHealth Innovation for Cancer Prevention Research Training Program Predoctoral Fellowship (Cancer Prevention and Research Institute of Texas grant RP210042). The content is solely the responsibility of the authors and does not necessarily represent the official views of the National Institutes of Health or the Cancer Prevention and Research Institute of Texas. Mervat Alharbi gratefully acknowledges King Faisal Specialist Hospital and Research Center in Saudi Arabia for providing partial financial support through a PhD scholarship.

## Abstract

**Background:** Establishing reliable reference dosimetry for ultra-high dose-rate (UHDR) beams (≥ 40 Gy/s) is challenging with conventional reference ionization chambers (IC) due to saturation effects arising from ion recombination. The Exradin A30 IC was introduced to overcome these challenges by utilizing an ultra-thin electrode spacing of 0.3 mm to enhance charge-collection efficiency (CCE).

**Purpose:** The aim of this study is to investigate the commercial A30 IC as a reference dosimeter for UHDR electron beams. It explores the dosimetric properties of the A30 IC, including leakage current, CCE, polarity, and beam quality correction factors.

**Methods:** Several measurements at the IntraOp® Mobetron® were acquired at a 1.5 cm distance from a medical UHDR electron accelerator head to achieve a maximum dose per pulse (DPP) of 9 Gy, and the instantaneous (or intra-pulse) dose rate (IDR) was 2.25 MGy/s using a 9 MeV electron beam. The measurements were obtained in a grounded water-equivalent plastic phantom, distilled water, and saline water. DPP values were adjusted by varying the SSD at a fixed pulse width (4 μs) while the pulse repetition frequency (PRF) was varied between 5 and 90 Hz. The CCE was calculated using EBT-XD radiochromic films under both UHDR and conventional beam conditions at identical dose and energy settings. Both CCE and $P_{pol}$ were also measured as a function of DPP and PRF in distilled and saline water. Beam quality correction factors, $k_Q$, were calculated using Monte Carlo and measured in electron beams from a TrueBeam linear accelerator.

**Results:** The A30 IC exhibited a leakage current of less than 2 fA. Both $P_{pol}$ and CCE decreased with increasing DPP, with CCE remaining in the 90-99% range across all three phantoms while polarity corrections decreased from 0.99 to 0.981 in both liquid and virtual water, respectively. Both CCE and $P_{pol}$ were observed to be independent of the PRF, ranging from 5-90 Hz, in distilled and saline water. Measured $k_Q$ values agreed with calculation to within 0.8% for all energies except 9 MeV, which showed a 2% discrepancy. **Conclusions:** The commercial A30 ionization chamber exhibited 5% recombination with DPP of up to 5 Gy in both distilled and saline water. The IC signal in a solid phantom was found to be dependent on charge buildup effects, which can be mitigated by utilizing grounded phantoms. When appropriate CCE corrections are applied, the A30 IC is a suitable reference dosimeter for UHDR electron beams.

## Introduction

Ultra-high dose rate (UHDR) has recently been introduced as a novel advancement in radiotherapy. UHDR radiotherapy is commonly defined as utilizing a radiation beam with a mean dose rate that exceeds 40Gy/s. The clinical and biological formalization of modern UHDR was formally established by Favaudon et al. (2014), who reported that UHDR irradiation achieves equivalent tumor control while sparing normal tissue relative to mice treated with identical doses but with conventional dose rate (CONV) RT, thereby widening the therapeutic window.[1] This preferential normal tissue sparing and isoeffective tumor control is called the FLASH effect. To facilitate the clinical translation of FLASH RT, the challenges related to accurate dosimetry and real-time beam monitoring must be addressed. Existing dosimeters that are heavily utilized in FLASH RT are primarily passive dosimeters such as alanine, thermoluminescent dosimeters (TLDs), optically stimulated luminescent dosimeters (OSLDs), and radiochromic films due to their dose-rate independence. Despite their utility, passive dosimeters have substantial drawbacks in their characteristics that limit their clinical utility, such as their poor temporal resolution, delayed read-out, and large uncertainties compared with ionization (ion) chambers (IC) that are routinely used in clinical practice. Alternative dosimeters that have been utilized in FLASH RT to measure the dose in real-time include diamond detectors, plastic scintillators, beam current transformers (BCTs), and Faraday cups. However, each of these dosimeters has its own respective limitations and uncertainties in UHDR RT and are not routinely used in current clinical practice.[2–4] To facilitate the clinical translation of FLASH RT properly, it is important to establish ion chambers for dose measurements, since they are the only type of detector recommended for use in national and international reference dosimetry protocols and guidelines.

The foundation of UHDR dosimetry for ion chambers was initially established by Boag and Wilson (1952) through their investigation of ion recombination and saturation effects in parallel-plate ionization chambers under high ionization-intensity conditions.[5] The high charge density generated by a single radiation pulse produces a cloud of residual positive space charge because electrons have much higher mobility than the positive charge carriers. This space-charge–induced electric field distorts and can significantly suppress the externally applied field in a conventional reference ionization chamber, thereby reducing the net electric field, slowing charge transport, and leading to severe recombination and saturation effects.[6,7] The source of the severe recombination and saturation effects that are commonly observed in ion chambers irradiated in UHDR beamlines (specifically electrons) arises from the dose per pulse (DPP) settings that are commonly used (> 0.5 Gy/pulse), indicating that it is the intensity of the radiation delivered that affects the chamber response.[8] In contrast, these effects can be independent of the pulse repetition frequency (PRF) or repetition rate because the charge collection time of the chamber is significantly shorter than the interval between pulses for most linear accelerators. Several strategies can be used to address these limitations. One approach is to increase the applied electric field, which enhances the drift velocity of charge carriers; however, excessively high fields may lead to charge multiplication, thereby compromising dosimetric accuracy. An alternative approach is to modify the ionization chamber geometry by reducing the electrode spacing. This approach is particularly effective because reducing the gap reduces the number of charge-carrier collisions before charge reaches the collector at the same electric field strength. This reduces the probability of recombination and charge multiplication compared with a larger-gap chamber. However, to mitigate saturation effects at DPP values typically used in preclinical studies, the electrode gap must be less than a millimeter in parallel-plate chambers, as experimentally and computationally validated in previous studies.[7,9]

The Exradin A30 (Standard Imaging, Middleton, WI) is an ultra-thin parallel-plate ion chamber with a 0.3 mm nominal electrode spacing to enhance charge-collection efficiency (CCE) (Figure 1). In its first study, the Exradin A30 was developed as a prototype parallel plate ion chamber that modified the existing design of the Exradin A10 by shrinking the electrode gap from 2 mm to 0.3 mm[8]. The current investigation uses the commercialized design of the Exradin A30, which is a waterproof enhancement of the original prototype, and evaluates it as a reference dosimeter for UHDR pulsed electron beams. Key characteristics including leakage current, CCE, polarity correction factor ($P_{pol}$), and beam quality correction factors are presented.

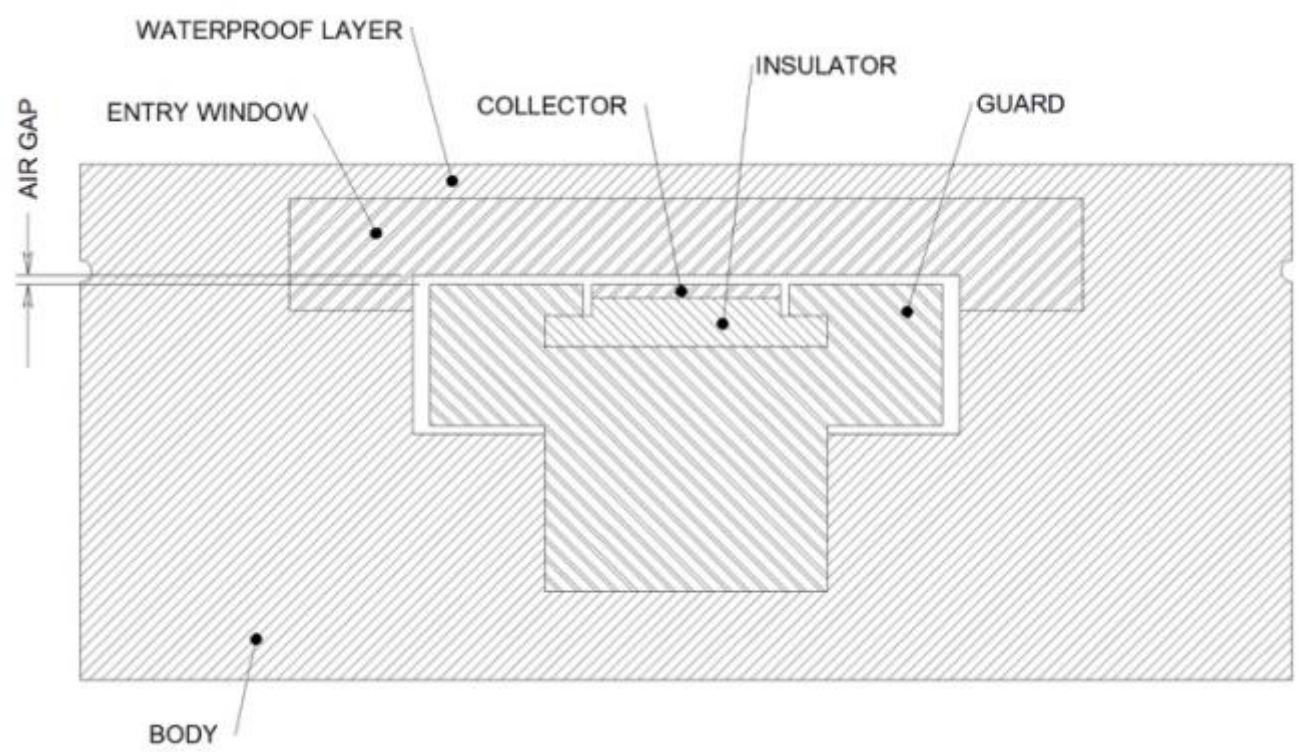

**Figure 1**: Schematic of the A30 IC design.

## Method

The IntraOp® Mobetron® is a linear accelerator that operates in both CONV and UHDR modes.[10] IC measurements were obtained in various grounded phantoms, including water-equivalent solid phantoms and a 1D water tank, which was initially filled with distilled water and later with saline (5 g/L). A Beam Current Transformer (BCT), integrated into the head of the Mobetron to monitor beam current, is shielded by a grounded Faraday cage to suppress electromagnetic interference, enabling stable and accurate real-time dose monitoring under UHDR conditions and demonstrating robustness to output fluctuations.[8,11] A depth of 1.5 cm was selected as the reference depth in this work based on the percentage depth dose (PDD) of the 9 MeV beam (4 μs, 30 Hz). At this depth, the PDD was relatively flat, thereby minimizing positional uncertainties and the associated dose variations. The BCTs were calibrated against radiochromic film positioned at a depth of 1.5 cm, and their readings were subsequently used for beam monitoring. The chamber was positioned at a physical depth of 1.2 cm to account for the 3 mm entrance window, yielding an effective measurement depth of 1.5 cm, consistent with AAPM TG-51 guidelines for reference dosimetry for a 9 MeV electron beam.

A 10-cm water-equivalent backscatter phantom was used, and a thin slab (0.3 cm) was positioned beneath the chamber to support the placement of a 2 mm aluminum sheet underneath it, which was connected to a grounding cable attached to the head of the Mobetron (Figure 2). The phantom was grounded to prevent transient charge buildup in nonconductive materials exposed to the high doses of radiation commonly arising from repeated measurements performed in UHDR radiation beams.[12] The A30 was connected to a MAX4000 electrometer (Standard Imaging, Middleton, WI). An electric field strength of 1000 V/mm (+ 300 V bias voltage) was applied, as this voltage was optimal to reduce the chamber's dependence on instantaneous dose rate, prevent ion multiplication, mitigate ion recombination, and improve the CCE.[8]

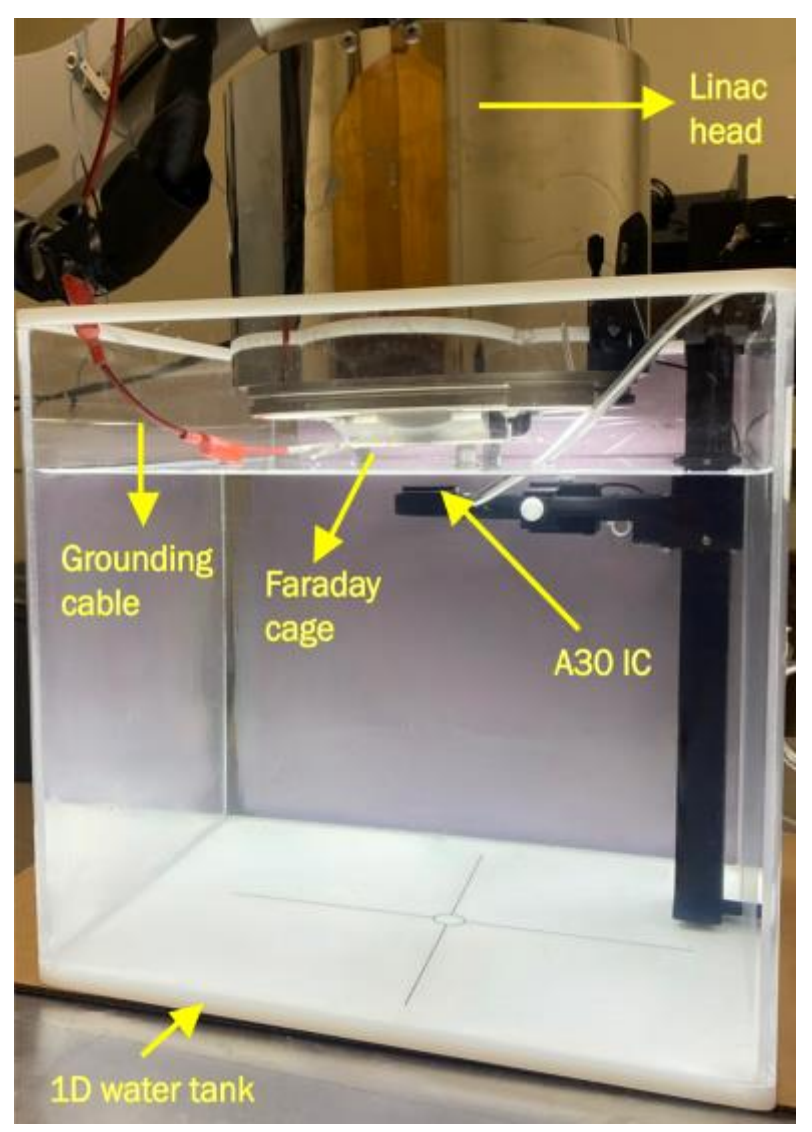

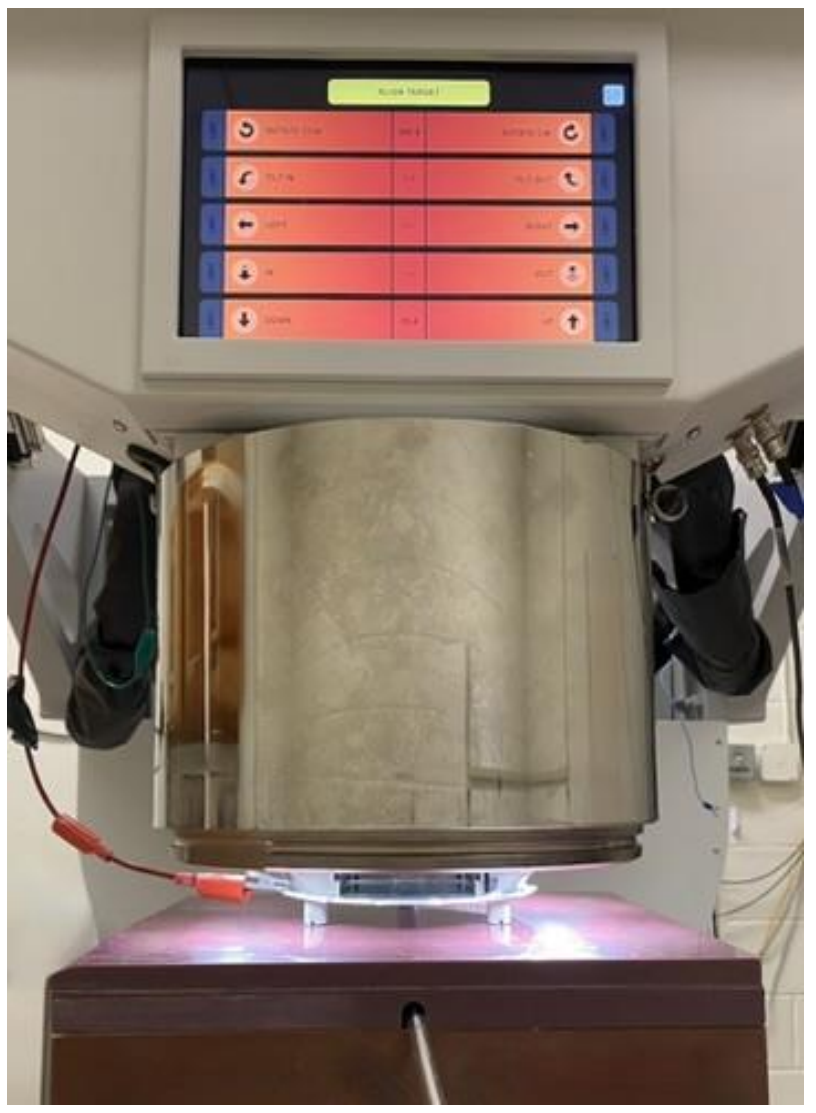

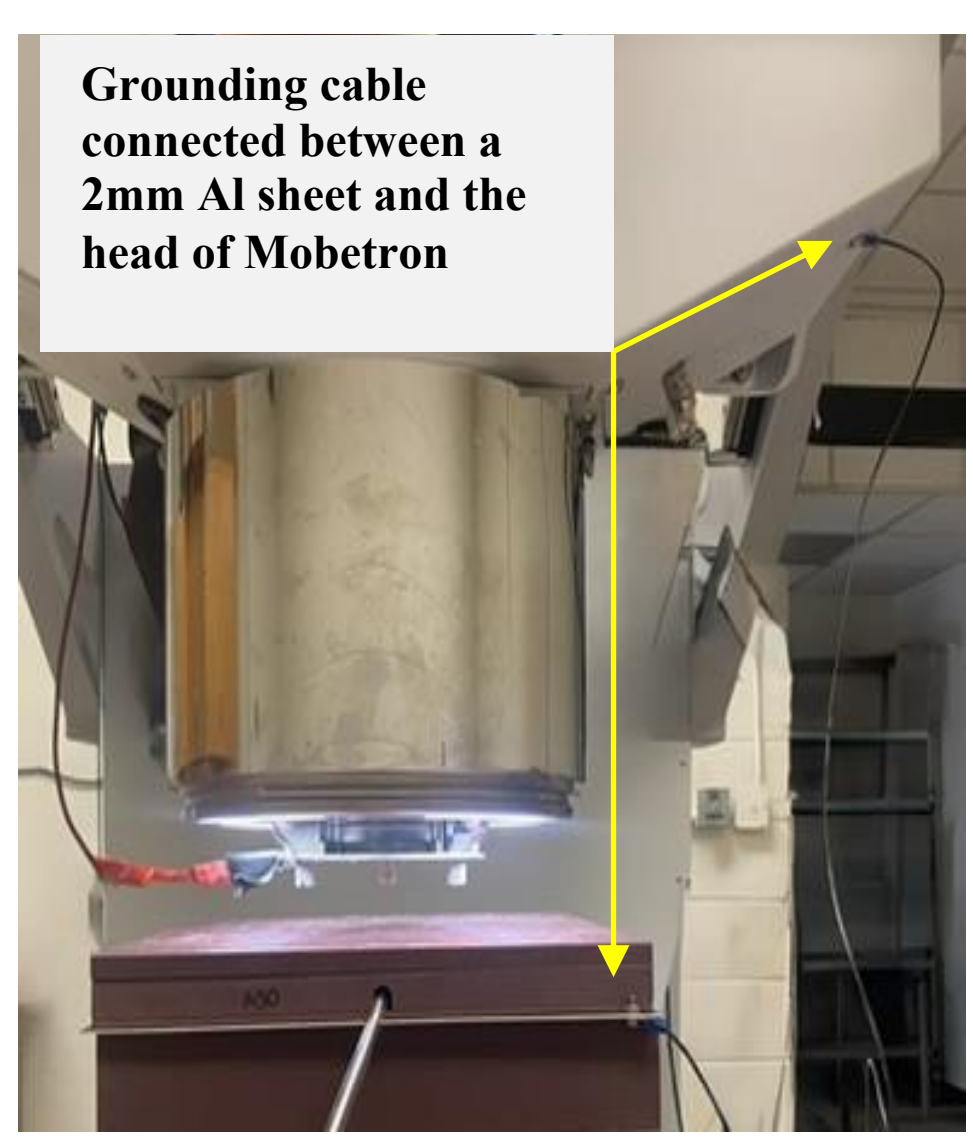

**Figure 2**: Measurement setup in a 1D water tank (left), virtual water (center), and grounded virtual water (right).

All measurements were performed three times at a 1.5 cm air gap between the Faraday cage attached to the exit window of the Mobetron head and the phantom surface, achieving a maximum dose per pulse (DPP) of 9 Gy (2.25 MGy/s instantaneous dose rate) with a 9 MeV electron beam. PW and SSD were adjusted independently to distinguish the effect of the pulse duration from those associated with DPP and IDR. Because DPP and PW were linearly correlated at a given SSD, the SSD was varied to modulate both DPP and IDR while maintaining a constant pulse duration of 4 μs. PRF was varied between 5-90 Hz. For each irradiation, measurements were normalized by the shielded BCT signal to correct for beam output fluctuations.[11]

The UHDR recombination effect was quantified as charge-collection efficiency (CCE) by comparing ionization chamber response under both UHDR and conventional beam conditions at identical dose and energy settings. These readings were normalized to doses measured with EBT-XD radiochromic film, which was selected for its demonstrated dose-rate-independent response within the UHDR range used in this study.[13] The CCE was calculated using the following equation [8]:

$$CCE = 100\% * \frac{\left(\frac{M_{raw} * P_{TP} * P_{pol}}{D_{film}}\right)_{UHDR}}{\left(\frac{M_{raw} * P_{TP} * P_{pol} * P_{ion}}{D_{film}}\right)_{Conv}}$$

$M_{raw}$ represents the uncorrected ion chamber reading; $P_{TP}$ is the correction factor for environmental conditions (temperature and pressure); $P_{ion}$ is the ion recombination correction factor for the conventional beam; and $D_{film}$ corresponds to the film dose. The polarity effect, $P_{pol}$, for the ion chamber was calculated by using the two-polarity voltage equation[14]:

$$P_{pol} = \left| \frac{(M_{raw}^{+} + M_{raw}^{-})}{(2 * M_{raw})} \right|$$

$M_{raw}^{+}$ and $M_{raw}^{-}$ represent the raw ion chamber readings obtained at opposite polarities. The A30 IC beam quality correction factors, $k_Q$, were measured on a Varian TrueBeam linear accelerator (Varian Medical Systems) using electron beams and cross-calibrated against a calibrated cylindrical chamber (Exradin A12). For each electron energy, measurements were performed at the reference depth, $d_{ref}$, and were corrected for the electrometer response, polarity, temperature, and pressure effects. Since the measured dose is equal for both chambers, $k_Q^{A_{30}}$ can be calculated as follows[14,15]:

$$(k_Q * M_{corr} * N_{D,W}^{60_{Co}})^{A_{30}} = (\mathrm{k}_{\mathrm{Q_{ecal}}} * k_Q' * M_{corr} * N_{D,W}^{60_{Co}})^{A_{12}}$$

$$(k_Q)^{A_{30}} = \frac{(\mathrm{k}_{\mathrm{Q_{ecal}}} * k_Q' * M_{corr} * N_{D,W}^{60_{Co}})^{A_{12}}}{(M_{corr} * N_{D,W}^{60_{Co}})^{A_{30}}}$$

**Beam quality correction factor simulations**

All Monte Carlo (MC) simulations in this work were executed in sequential mode using 2000 concurrent independent simulation jobs (2000 cores) on the University of Wisconsin-Madison Center of High Throughput Computing (CHTC) cluster.

Beam quality correction factors, $k_Q$, were calculated using the TOol for PArticle Simulation (TOPAS), which is a GEometry ANd Tracking (GEANT4) wrapper.[16,17] $k_Q$ for electron beams with $^{60}Co$ as the reference beam quality can be calculated by:

$$k_Q = \frac{f_Q}{f_{Q_0}} = \frac{\left(\frac{D_w}{D_{IC}}\right)_Q}{\left(\frac{D_w}{D_{IC}}\right)_{Q_0}}$$

where $D_w$ is absorbed dose to water, $D_{IC}$ is the absorbed dose to the ion chamber, $Q_0$ is the reference beam quality of $^{60}Co$, and $Q$ is the clinical beam quality.[18,19] MC parameters used in this study are provided in Table 1. A 30×30×30 $cm^3$ water phantom was simulated at a 100 cm source-to-surface distance (SSD) and percent depth dose (PDD) was scored in voxels of 2×2×0.5 $mm^3$ with 0.5 mm being along the depth direction. These simulations provide the $D_W$ values for a given beam quality. The A30 ion chamber was modeled using the vendor-provided drawings and information on the utilized materials for construction. As described in our previous work, the $f_{Q_0}$ factor was calculated with the ion chamber placed at 5 cm depth using a standard $^{60}Co$ photon spectrum collimated to create a 10x10 $cm^2$ square field size at 100 cm SSD.[18,20] For electron beams, phase space files, proximal to the jaws, provided by the vendor of the Varian TrueBeam linear accelerator were employed with nominal energies of 6, 9, 12, 15, 18, 20, and 22 MeV.[21] Range shifters, composed of 3-5 mm water-equivalent thicknesses, were employed to achieve intermediate beam qualities between the nominal energies, thereby generating a more densely populated dataset. Physical jaws and electron applicator for the reference 10×10 $cm^2$ field size were explicitly modeled using the manufacturer-provided drawings. The beam quality metric, $R_{50}$, was calculated using the PDDs as the depth where the absorbed dose drops to 50% of its maximum value, which was then used to determine the reference depths using the relationship provided in the AAPM TG-51 and WGTG51 Report 385 protocols[14,22,23]:

$$d_{ref} = 0.6R_{50} - 0.1cm$$

For each beam quality, the front face of the air cavity of the ion chamber was placed at the $d_{ref}$ and dose was scored to the air cavity determining $D_{IC}$ . Following the WGTG51 Report 385 protocol for electron beams, the $k_{Q_{ecal}}$ value was reported as the $k_Q$ for the $R_{50} = 7.5cm$ beam quality and $k'_Q$ was calculated by normalizing the $k_Q$ to the $k_{Q_{ecal}}$ value. Beam quality correction factors were fit to a function of the following form:

$$k'_Q = a_1 + \left(b_1 \times e^{-R_{50}/c_1}\right)$$

where $a_1$, $b_1$, and $c_1$ are fitting parameters.

**Table 1**: Monte Carlo simulation parameters used in this work in accordance with TG-268.

| Item name | Description | References |
|---|---|---|
| Code version | TOPAS v3.7 | [16,17] |
| Validation | Vendor-provided dataset | [21] |
| Timing | ~12 hours per job | |
| Source description | $^{60}$Co: photon energy spectrum<br>Electrons: phase space files with an optional range shifter | [20] |
| Transport parameters | G4EMStandardOpt4 | |
| Variance reduction | None | |
| Scored quantity | Absorbed dose to medium | |
| Statistical uncertainty | <0.50% | |
| Postprocessing | None | |

## Results

The A30 IC leakage was evaluated prior to all measurements and was found to be < 2 fA. The $P_{pol}$ was measured to be 1.0009 under Co-60 reference calibration conditions at UWADCL.

DPP values between 1.9 and 9 Gy were obtained by adjusting the SSD using the setup and phantoms shown in Figure 2. As shown in Figure 3, the CCE decreased monotonically with increasing DPP across all media, dropping from approximately 99% to 90% in both liquid and Virtual Water™ (Standard Imaging, Middleton, WI). The difference in $P_{pol}$ between distilled, saline, and virtual water is less than 1%, and values range from 0.98 to 0.99 across all DPP and pulse widths. Figure 4 illustrates the difference in CCE and $P_{pol}$ between grounded and ungrounded virtual water. The CCE and $P_{pol}$ were approximately 2% lower in ungrounded virtual water than in grounded virtual water across all DPP values.

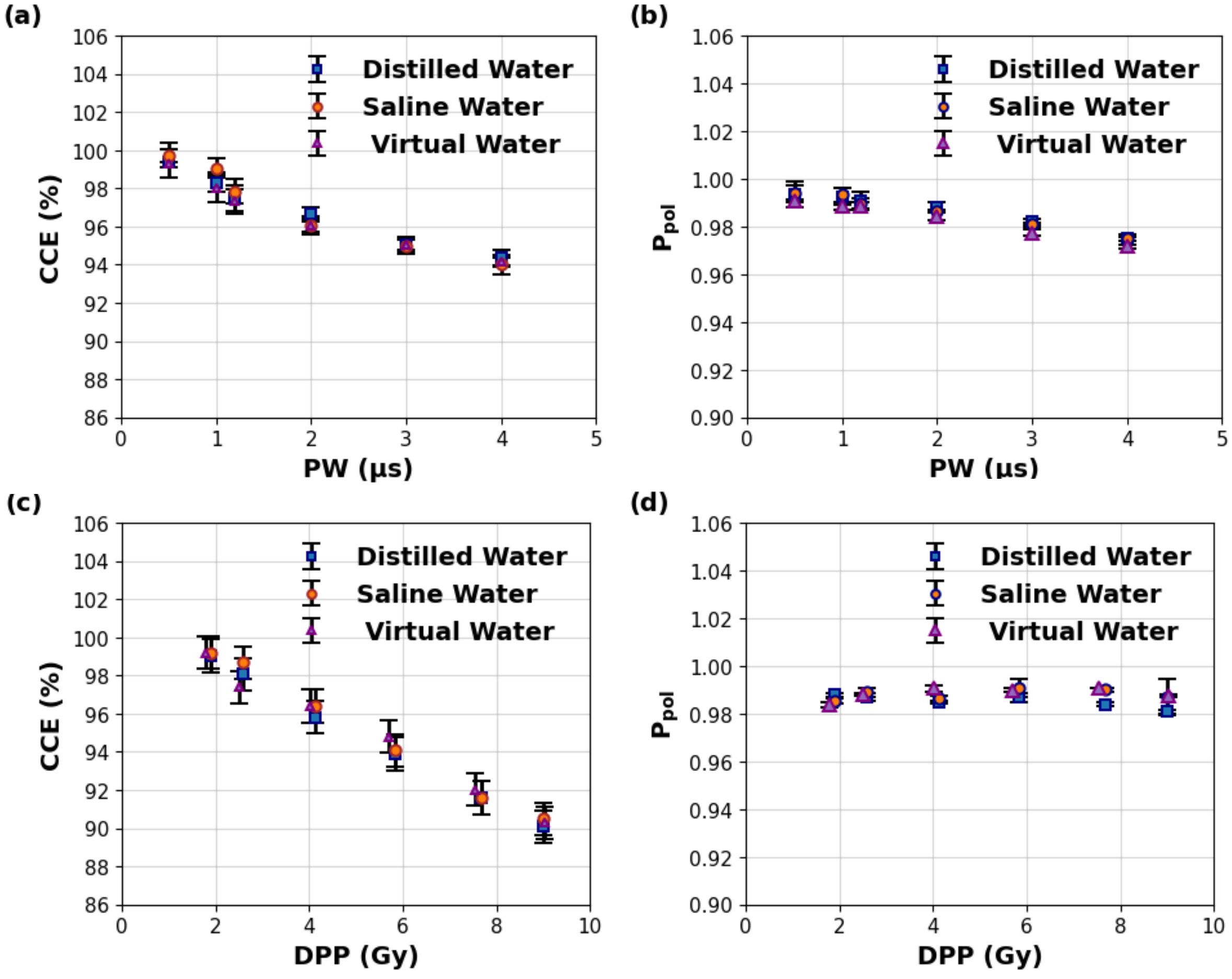

**Figure 3:** Charge collection efficiency (CCE) and polarity correction factor ($P_{pol}$) as a function of (a, b) pulse width (PW) and (c, d) dose per pulse (DPP) in distilled, saline, and virtual water. The error bars indicate one standard deviation of the measured values of each data point.

No PRF dependence was observed for either CCE or $P_{pol}$ over the investigated range of 5-90 Hz in both distilled and saline water (Figure 5). The CCE values measured in the A30 IC as a function of pulse repetition frequency (PRF), ranging from 5-90 Hz, were shown to be constant at a given DPP of 9.6 Gy, with CCE maintained at approximately 93.8% and 94.8%.

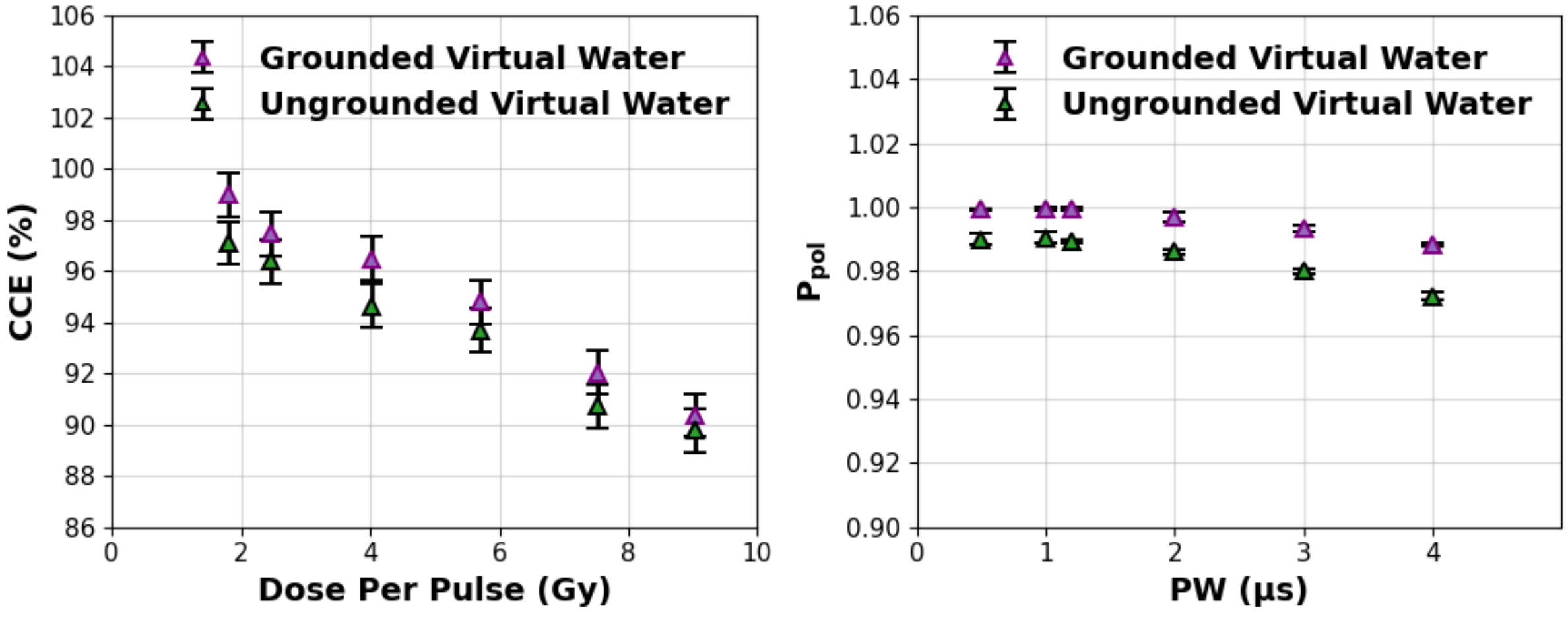

**Figure 4**: Charge collection efficiency (CCE) (left) and polarity correction factor ($P_{pol}$) (right) in a grounded and ungrounded virtual water phantom.

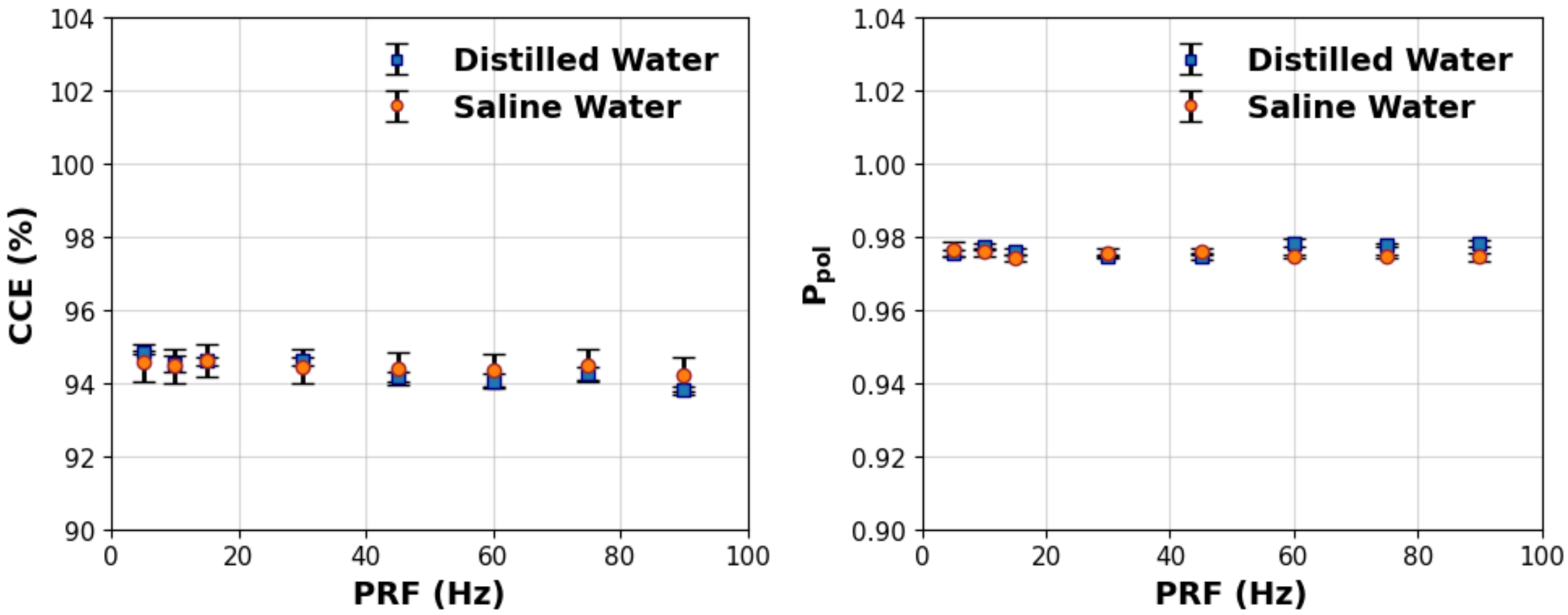

**Figure 5:** Charge collection efficiency (CCE) (left) and polarity correction factor ($P_{pol}$) (right) as a function of pulse repetition frequency varied from 5- 90 Hz, corresponding to mean dose rates of 69- 1242 Gy/s, and at 4 µs in distilled and saline water. The error bars indicate one standard deviation of the measured values of each data point.

**Beam quality correction factor**

Figure 6 shows the PDDs for each electron beam simulated in this work. The nominal and range shifted energies had $R_{50}$ values ranging from 2.3-8.3 cm. To capture the steeper dependence of $k_Q$ on beam quality at lower electron energies, we intentionally sampled a denser set of beam qualities in this region, whereas fewer beam qualities were required at higher energies where the $k_Q$ curve is flatter.

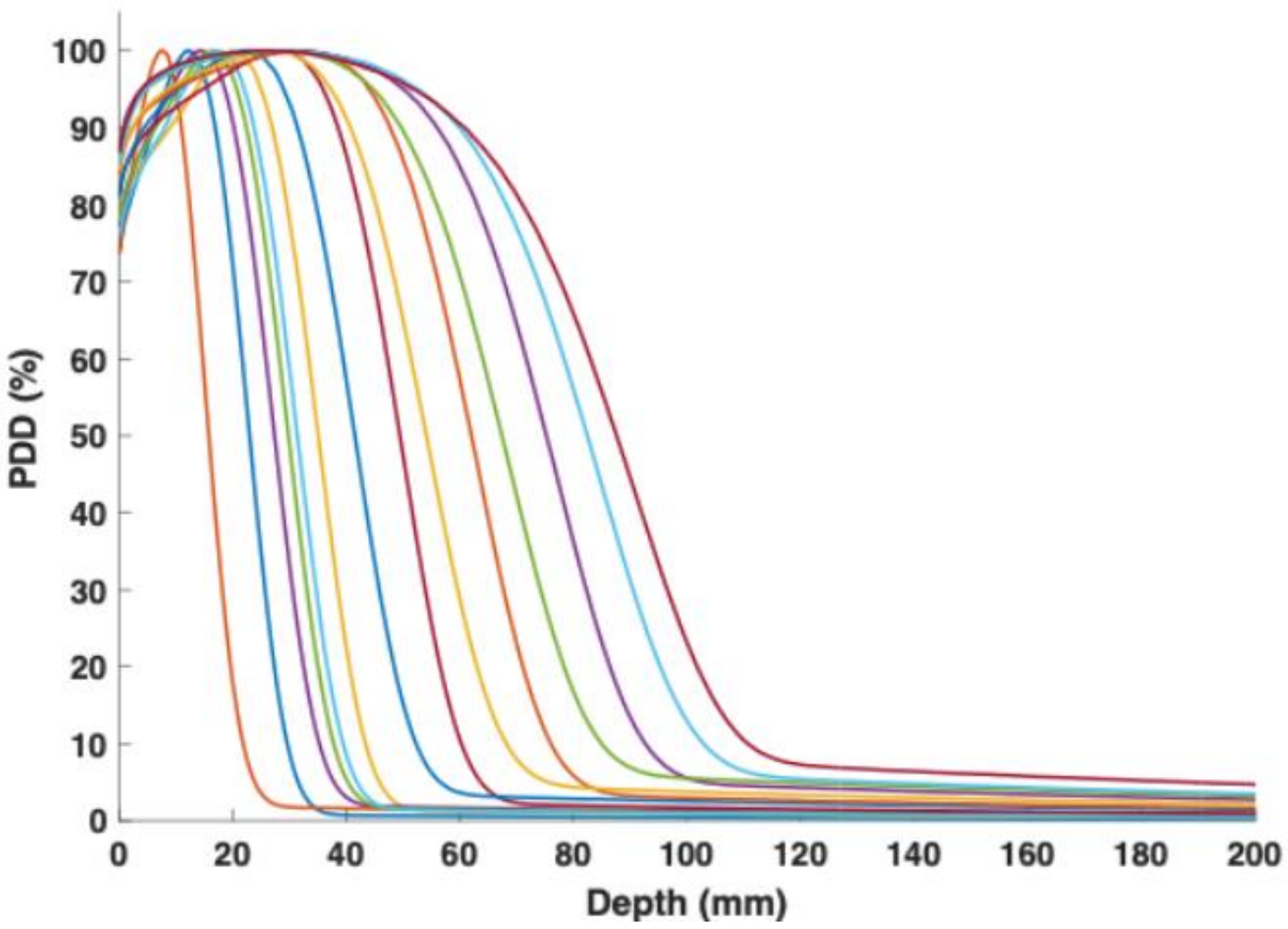

**Figure 6**: The percent depth dose (PDD) curves for the electron beams employed in $k_Q$ calculations.

Figure 7 displays the A30 beam quality correction factors in the form of $k_Q$ and $k'_{\mathrm{Q}}$. The $k_{Q_{ecal}}$ was calculated to be 0.915 ± 0.002. The chosen function fit well to the data with root mean square error (RMSE) of 0.0014. As expected, the beam quality correction deviation from unity is larger for higher energy electron beams due to the decrease in water-to-air stopping power ratios, $\left(\frac{S_w}{S_{air}}\right)$, with increasing electron energies. The ion chamber response is sensitive to the employed beam quality and corrections must be used when measuring absorbed dose to water in a clinical beam.

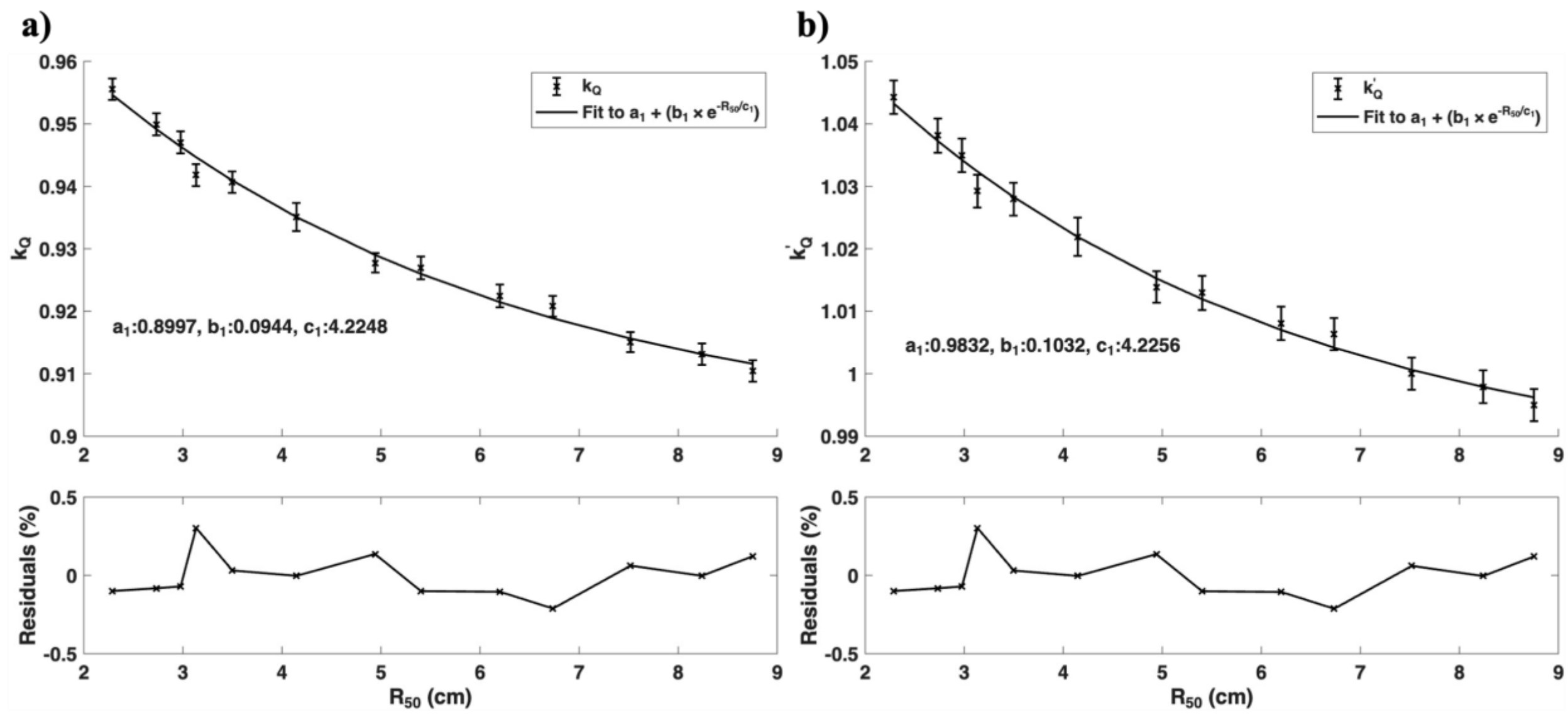

**Figure 7**: a) $k_Q$ correction factor and b) $k'_Q$ correction factor for the A30 ion chamber as a function of $R_{50}$. The fitting parameters and residuals have been reported. The error bars indicate the Type A uncertainty at σ = 1 in the calculated values.

The $k_Q$ response of the A30 ion chamber was measured experimentally and compared with MC simulation for conventional electron beams from two linear accelerators (Linac 1 and Linac 2) at various nominal energies (Figure 8). In linac 1, the $k_Q$ is determined as the mean of three independent measurements obtained using three A30 ICs, while in linac 2, the $k_Q$ was measured from a single IC. The agreement between the two methods remained below 0.8% across all electron energies.

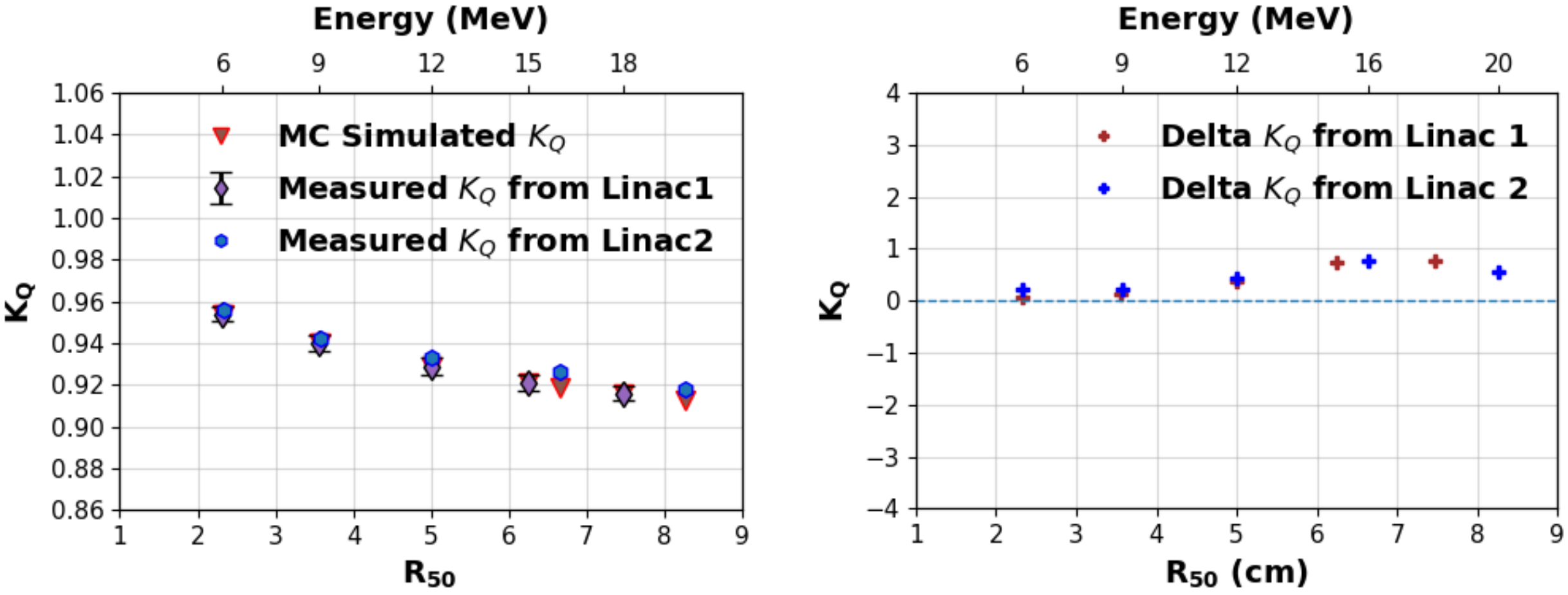

**Figure 8:** MC-simulated and measured beam quality correction factor ($k_Q$) from two linear accelerators under conventional beam conditions across different electron energies (left), and the percent difference between the simulated and calculated $k_Q$ (right). The error bars represent the standard error of the mean of three measurements obtained from three independent ion chambers.

## Discussion

This work provides a comprehensive characterization of the commercial Exradin A30 under pulsed UHDR electron beam conditions relevant to electron FLASH, including leakage, polarity behavior, and CCE. While the prototype A30 design demonstrated that sub-mm electrode spacing and high electric field strength can substantially mitigate saturation effects in UHDR RT[8], the commercial A30 adds practical robustness through a waterproof configuration and is therefore more suitable for routine measurements in both liquid water and solid water phantoms. Our results show that, when operated at 1000 V/mm, the chamber demonstrates low leakage, stable polarity behavior in grounded water-based media, and a CCE that remains >90% at DPP up to 5 Gy (1.25 MGy/s IDR), supporting the feasibility of the commercial A30 as a candidate reference dosimeter for UHDR beamlines.

Across all tested media, the A30 CCE decreased with DPP. This behavior is consistent with established physical mechanisms in which high ionization-intensity pulses produce residual space charge that distorts and suppresses the effective electric field, thereby increasing recombination losses and reducing the CCE. The magnitude of the observed reduction in CCE with DPP aligns with prior findings that instantaneous dose rate is the dominant driver of ion chamber saturation in electron UHDR beams, while other beam parameters, such as PRF, have little to no influence when charge is fully collected before the next pulse.[7,8,24] Practically, these findings reinforce that reference-class use of the A30 in UHDR electron beams should be implemented with an explicit and validated CCE correction framework as a function of DPP, rather than relying on conventional $P_{ion}$ corrections that were developed for conventional dose rates and do not fully capture UHDR conditions. One important thing to note was the difference in the magnitude of the CCE and $P_{pol}$ as a function of DPP between grounded and ungrounded virtual water. It is likely that the culprit for the discrepancies between the CCE and $P_{ion}$ arises from charge buildup in the virtual water phantom after repeated irradiations at UHDR[12]. This effect was observed to be more pronounced in virtual water than in liquid water due to its low electrical conductivity, which limits charge dissipation and can lead to charge buildup within the irradiated material, potentially inducing an electric field.[12]

No measurable PRF dependence was observed in CCE or $P_{pol}$ over 5–90 Hz in distilled and saline water. This supports the hypothesis that the dominant recombination and saturation losses occur intra-pulse, and charge collection is sufficiently complete between pulses under the investigated conditions. This is consistent with previously reported observations that PRF, which modifies the mean dose rate, is less influential than DPP and instantaneous dose rate in pulsed electron UHDR beams for many ion chamber geometries.[8,7,24] However, it is worth taking into consideration that should linear accelerators be developed for UHDR RT that operate with PRFs in the kHz or MHz range, then it would be worth reinvestigating PRF dependence as the time between pulses may be comparable to the ion collection times. From a workflow standpoint, the absence of PRF dependence reduces the complexity for real-time beam monitoring from an electrometer standpoint, since chamber response can be parameterized primarily by DPP (and, if needed, pulse width at low field strengths), rather than requiring PRF-specific corrections.

In grounded liquid water conditions (distilled and saline), $P_{pol}$ remained close to unity with modest variation across DPP, whereas slightly lower $P_{pol}$ values were observed in grounded virtual water. This medium dependence strongly suggests that polarity behavior in UHDR measurements is not solely an intrinsic chamber property but is also influenced by the electrical boundary conditions and charge transport in the surrounding medium and measurement setup. In UHDR electron beams, charge deposition in the surrounding medium and at material interfaces can generate charge buildup and field perturbations that differ between solid phantoms and conductive liquids. Such effects can manifest polarity differences and apparent signal shifts unrelated to the true chamber collection efficiency. While both distilled and saline water were grounded in this study, the conductivity difference between these media did not translate into meaningful $P_{pol}$ differences, suggesting that grounding and overall boundary conditions, rather than ionic strength alone, dominate the observed behavior. These findings motivate a practical recommendation: for reference-class UHDR measurements with the commercial A30, well-controlled grounding in the experimental setup provides more stable polarity behavior than ungrounded configurations under the tested conditions. The measured A30 $k_Q$ values showed generally good agreement with Monte Carlo predictions across all energies, with deviations below 0.8%. These results highlight the importance of repeat measurements and platform consistency when establishing reference-class correction factors for a new chamber type, particularly when transferring protocols between clinical linacs and specialized UHDR beamlines.

Taken together, these data support the commercial A30 as a strong candidate for UHDR electron reference dosimetry, provided that its use is constrained to an experimentally validated dose rate range and accompanied with appropriate correction methodology. A practical implementation approach for using the A30 is: (1) operate at 1000 V/mm, (2) characterize and apply CCE versus DPP corrections for the specific beamline, grounded medium, and geometry, (3) use water-based mediums with robust grounding where possible to stabilize polarity behavior, and (4) avoid extrapolating corrections beyond the tested DPP and pulse width space without additional validation. Under these conditions, the A30's low leakage and stable performance in conductive liquid media support its use as a routine reference detector for calibration and commissioning workflows in electron FLASH beamlines.

This study has limitations that should be addressed in follow-up work. First, the working standard relied on has been a combination of film and BCT dose reporting for CCE calculations, which introduces uncertainty that can be reduced by additional independent standards (for example, calorimetry or alternative dose-rate independent dosimeters). Second, the observed medium dependence of $P_{pol}$ indicates that the chamber response can be sensitive to phantom material and grounding, motivating more systematic isolation of interface and transient-current effects in UHDR environments. Third, it has been established in a previous study[8] that the CCE is dependent on a combination of DPP and PW, where the CCE may decrease at a given DPP delivered at shorter PWs (thereby higher instantaneous dose rates). Simulated data has demonstrated that this effect can be mitigated at higher electric field strengths (but not so high as to induce charge multiplication), or smaller electrode gaps (e.g. 0.1 mm).[8] Finally, additional testing across beam energies, higher DPP values, and alternate electrometer systems would strengthen generalization and help quantify instrumentation contributions such as impulse response and saturation behavior in pulsed readout electronics.

## Conclusion

This study provides a comprehensive evaluation of the commercial Exradin A30 parallel-plate ion chamber for UHDR pulsed electron dosimetry and its potential use as a reference dosimeter in electron FLASH beamlines. The chamber exhibited negligible leakage current (~1 fA) and stable operation at 1000 V/mm. The A30 response showed a clear DPP dependence, with CCE decreasing as DPP increased, reaching approximately 90% in water-based media at the highest tested DPP (~9 Gy/pulse) and IDR (2.25 MGy/s). No measurable dependence on PRF was observed over 5–90 Hz. Polarity effects were minimal in grounded media but larger in ungrounded virtual water, indicating some sensitivity to measurement environment (likely due to charge buildup). Measured $k_Q$ values agreed with Monte Carlo predictions within <0.8% for all electron energies. Overall, the commercial A30 is well-suited for UHDR electron reference dosimetry, provided that CCE corrections as a function of DPP are applied.